\documentclass[12pt,a4paper]{article}

\usepackage{amsmath,epsfig,amssymb}

\begin{document}
 \bibliographystyle{unsrt}

\begin{center}
{\Large \bf Surprises in Diffuse Scattering}

 \end{center}
 \vspace{3mm}

 \begin{center}  {\sc Moritz H\"offe} and {\sc Michael Baake} 
 
\vspace{5mm}

 Institut f\"ur Theoretische Physik, Universit\"at T\"ubingen, \\

 Auf der Morgenstelle 14, 72076 T\"ubingen, Germany

\end{center}


\vspace{10mm}
\begin{abstract} 
Diffuse scattering is usually associated with some disorder in
the analyzed material. Different kinds of disorder may produce different diffuse
scattering -- or not. In this letter, we demonstrate some aspects of the variety of diffuse
scattering that occurs even in very simple examples, and how unawareness may
lead astray.
\end{abstract}

\parindent15pt
\vspace{5mm}

\section{Introduction}

For a long time, the diffuse part in diffraction spectra played only a minor role in
crystallography. This was mainly due to experimental restrictions but also to a lack
of theoretical studies. In this letter, we investigate some simple models with
disorder, both deterministic and random, where the
diffuse background of the diffraction must not be neglected in the structural analysis.

We first start with an introduction to the language of mathematical diffraction
theory, which is necessary to get a full understanding of the spectrum. The 
following two
examples have exactly the same diffraction pattern albeit their disorder is of
completely different type. We then discuss models with singular continuous
diffraction, a type of spectrum that is not present in classical crystallography but 
may appear in random structures with long range order. This is followed by the Ising 
lattice gas where the Bragg spectrum alone does not reflect the correct symmetry. Last, we compare
square ice with an alternative model that can only be distinguished by the
diffuse scattering.

\section{Diffraction theory}

Let us recall some notions of diffraction theory \cite{Cowley,BH}, adapted to our 
purposes using the measure theoretic approach \cite{Hof95}. On the assumption of
kinematical diffraction in the Fraunhofer picture, the diffracted intensity per atom
$\widehat{\gamma}_{\omega}$ is given by the Fourier transform of the autocorrelation
$\gamma_{\omega}$. Let the unit point measure (Dirac distribution) $\delta_x$ idealize an atom 
at position $x$ and $\Lambda$ denote the point set of all possible atomic positions.
Then the atomic structure is described by the so-called weighted Dirac comb
\begin{equation}
\omega_{\Lambda}=\sum_{x\in\Lambda} w(x) \delta_x,
\end{equation}
where $w(x)\in\{0,1\}$ indicates whether the point $x$ is actually occupied or not.
More generally, $w(x)$ can be a bounded complex weight function. The
autocorrelation (Patterson function) $\gamma_{\omega}$, assuming it is well defined, 
is given by
\begin{equation}
\gamma_{\omega}=\sum_{z\in\Delta} \nu(z) \delta_z,
\end{equation}
with $\Delta=\Lambda-\Lambda$ being the set of difference vectors which we assume discrete
and closed. The autocorrelation
coefficient $\nu(z)$ can be calculated according to
\begin{equation}
   \nu(z) \; = \; \lim_{R\to\infty} \frac{1}{{\rm vol}(B_R)}
          \sum_{\stackrel{\scriptstyle y \in \Lambda_R}
                         {\scriptstyle z-y \in \Lambda}}
               w(y) \, \overline{w(z-y)} \, ,
\end{equation}
where $B_R$ is the ball of radius $R$ around the origin,
$\Lambda_R=\Lambda\cap B_R$ and \raisebox{1ex}{$\overline{\hphantom{x}}$}
complex conjugation.
If $w(x) \in \{0,1\}$, $\nu(z)$ is the frequency of having two scatterers
at distance $z$.

The Fourier transform of a function $\phi$ is
\begin{equation}
\widehat{\phi}(k)=\int_{\mathbb R^n} e^{-2 \pi i k\cdot x} \phi (x) \, dx,
\end{equation}
and we use the standard theory of tempered distributions from here, see \cite{BH} for 
our conventions.

The diffraction image, $\widehat{\gamma}_{\omega}$, is a positive measure that tells
us how much intensity is scattered into a given volume element. So, by the general
decomposition theorem \cite[Ch.\ I.4]{RS}, it admits a unique 
decomposition with respect to
Lebesgue's measure into three parts, $\widehat{\gamma}_{\omega} = 
(\widehat{\gamma}_{\omega})_{pp}+(\widehat{\gamma}_{\omega})_{sc}+
(\widehat{\gamma}_{\omega})_{ac}$. The pure point part $(\widehat{\gamma}_{\omega})_{pp}$
consists of the Bragg peaks, the absolutely continuous $(\widehat{\gamma}_{\omega})_{ac}$
corresponds to the usual diffuse background that can often be described as a continuous 
function, otherwise as an $L^1$-function.
The singular continuous component $(\widehat{\gamma}_{\omega})_{sc}$ lies somewhere in 
between, i.e.\ it is neither a smooth
function nor ``as singular'' as a Bragg peak, for a precise definition see \cite{RS}.

\section{Random versus deterministic disorder}

In the following, we present two one-dimensional examples with completely different type of
(dis)order but displaying the same diffraction spectrum, i.e.\ they are 
homometric.

The first model is a Bernoulli system on $\mathbb Z$. The structure is given by 
the stochastic Dirac comb
$\omega^{B}=\sum_{m\in\mathbb Z} \eta^{B}(m) \delta_m$ where $\eta^{B}(m)$ is a 
family of random variables that take the values $h_1$ and $h_2$ with
probabilities $p_1$ and $p_2$. For such a lattice gas, one can prove \cite{BH,BM} that 
its diffraction measure almost surely consists of a uniform pure point and a
constant absolutely continuous part,
\begin{equation}
\widehat{\gamma}_\omega^{B}=|\langle \boldsymbol h\rangle|^2 \sum_{m\in\mathbb Z} 
\delta_m +
\text{Var} ({\boldsymbol h} ),
\end{equation} 
with mean $\langle\boldsymbol h\rangle=p_1 h_1+p_2 h_2$ and variance 
$\text{Var}( \boldsymbol h ) 
 = \langle | \boldsymbol h|^2\rangle -|\langle\boldsymbol h\rangle|^2$. For
$h_1=1$, $h_2=0$ and $p_1=p_2=1/2$, we have $|\langle\boldsymbol h\rangle|^2 = 
\text{Var}( \boldsymbol h )=1/4$.

Let us compare this with the
Rudin-Shapiro sequence, where the atoms are distributed deterministically. We construct 
$\omega^{RS}=\sum_{m\in\mathbb Z} \eta^{RS}(m) \delta_m$ according to the rule
\begin{gather}
\eta^{RS}(0)=1, \quad \eta^{RS}(-1)=0, \nonumber\\
\eta^{RS}(4m)=\eta^{RS}(4m+1)=\eta^{RS}(m), \\
 \eta^{RS}(4m+l)=
\frac{1}{2}\left(1-(-1)^{m+l+\eta^{RS}(m)}\right), \; l=2,3  \nonumber
\end{gather}
Alternatively, this sequence may be defined via a substitution rule on a four letter
alphabet \cite{GL}. For the initiate, we add that we use the square of the
traditional substitution to get the fixed point of the bi-infinite
sequence. The diffraction spectrum can be calculated
rigorously \cite[pp.\ 165--168]{Quef}. With our choice of the scattering strenghts, 
this results in
\begin{equation}
\widehat{\gamma}_\omega^{RS}=\frac{1}{4} + \frac{1}{4}\sum_{m\in\mathbb Z} \delta_m.
\end{equation}
Amazingly, though the structure is completely deterministic, its two-point correlations
are destroyed systematically so that only a constant diffuse background remains in 
$\widehat{\gamma}_\omega$.

Therefore, with $h_1=1$ and $h_2=0$, we have $\widehat{\gamma}_\omega^{B}=
\widehat{\gamma}_\omega^{RS}$, a
distinction of the original structure on the basis of the diffraction spectrum 
is impossible.
Even for finite systems, the slight difference in the fluctuations of the background
is almost undetectable. Since the Fourier transform is unique, this means that the
Rudin-Shapiro sequence is homometric to the above Bernoulli chain (a statement which
is true with probability one). The remarkable feature of this example is the inclusion
of a diffuse background, see \cite{Zobetz} for other cases of homometry.

Note that this example explores the full entropy range: the Bernoulli case has
entropy $\log 2$, the
maximal value for a binary system, while Rudin-Shapiro has entropy 0. It is clear that
one can find other examples with the same diffraction image but entropy between these
extremal values. So, a resolution of the corresponding reconstruction problem, unless
extra information is available, needs an optimization approach, e.g.\ by choosing the
structure which maximizes the configuration entropy.

\section{Singular continuous spectra}

There is no simple characterization of the singular continuous part of a diffraction
spectrum, since this term covers the whole range (in the sense of tempered distributions)
between a continuous function, resp.\ $L^1$-function, and a Dirac distribution. 
As indication one may use the scaling behaviour \cite{GL,AGL}
of the intensities with the system size $N$: $\widehat{\gamma}_{pp}\sim N$,
$\widehat{\gamma}_{ac}\sim const.$ and $\widehat{\gamma}_{sc}\sim N^{\alpha}$,
$0<\alpha<1$, but this recipe is not correct in general \cite{Hof97}.
A classical example where the scaling argument yields the right answer is the Thue-Morse
sequence \cite{Quef,GL}.
Singular continuous spectra usually appear in structures where the constructive interference
that is responsible for the Bragg peaks is impaired by some randomness. Nonetheless,
some long range order is still strong enough to prevent the peaks from becoming completely
diffuse. Note that singular ``peaks'' can never be isolated -- a property that further
complicates their analysis.

As an illustration in one dimension, let us consider a ``tiling'' of the line without
gaps and overlaps which consists of two prototiles (intervals) of length 1 and
$\tau=(1+\sqrt{5})/2$. The unit scatterers are located on the left endpoints of the
intervals. By placing the intervals randomly with arbitrary (but fixed) probabilities, 
the resulting 
spectrum is absolutely continuous except for the trivial Bragg peak at $k = 0$, see  
\cite[Thm.\ 2]{BH}. On the other hand, arranging these intervals as a Fibonacci sequence 
leads to the well-known quasicrystal with purely discrete diffraction spectrum,
cf \cite{Hof95}. 

With the same tiles, one can now also construct a ``structure intermediate between 
quasiperiodic and random''. This is achieved using the so-called circle sequence
\cite{AGL}. The positions $x_n$ of the atoms are given by
\begin{equation}
x_n-x_{n-1}=1+\xi \mathbf 1_{[0,\beta)}(n\alpha)
\end{equation}
where $0<\beta<1$, $\mathbf 1_{[0,\beta)}$ is the characteristic function of
the interval $[0,\beta)$ and $x_0$,
$\xi \in\mathbb R$ are parameters; assume $x_0=0$ for simplicity. In
order to have the same tile lengths and density $d=1/(1+\beta\xi)$ as in the Fibonacci
case, we choose $\xi=\tau-1$ and $\beta=2-\tau$. As was shown by Hof \cite{Hof96}, the
circle sequence has purely singular continuous diffraction spectrum apart from the trivial
Bragg peak for every irrational $\xi$, every $\beta$ and generic (``most'') $\alpha$.
Nevertheless, the assessment in each explicit case is open.

In two dimensions, a straightforward generalization of the above mentioned Thue-Morse
chain is given by the next example. It is constructed by using a two-dimensional
substitution rule \cite{Allouche,Hermiss} on a two-letter alphabet:
\begin{equation}
\rho: \, 
a \; \mapsto \; \begin{matrix} b & a \\ a & b \end{matrix}\; ; \quad
b \; \mapsto \; \begin{matrix} a & b \\ b & a \end{matrix}\; ,
\end{equation}
or, using two elementary cells with different atoms, 
\begin{figure}[ht]
\centering\epsfig{file=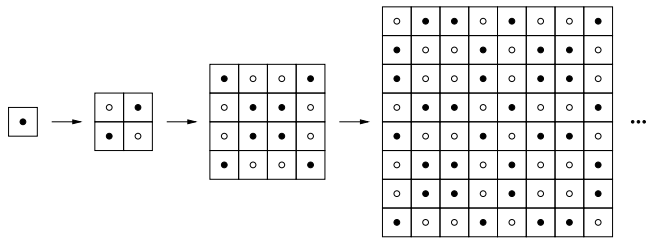,width=8.5cm}
\caption{Construction of the 2D Thue-Morse pattern.}
\label{tm2dstruct}
\end{figure}

\begin{figure}[ht]
\centerline{\epsfig{file=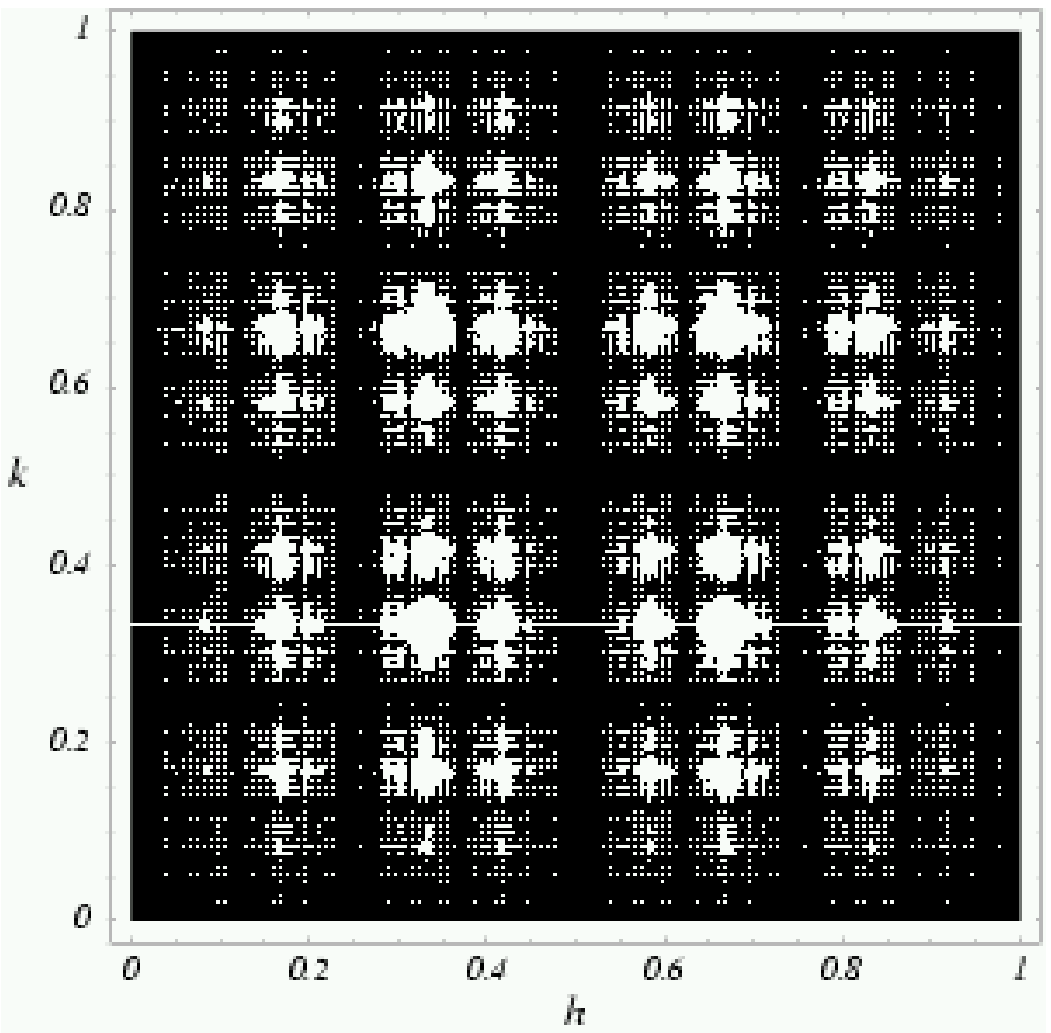,width=7cm}}
\vspace{.3cm}
\centerline{\epsfig{file=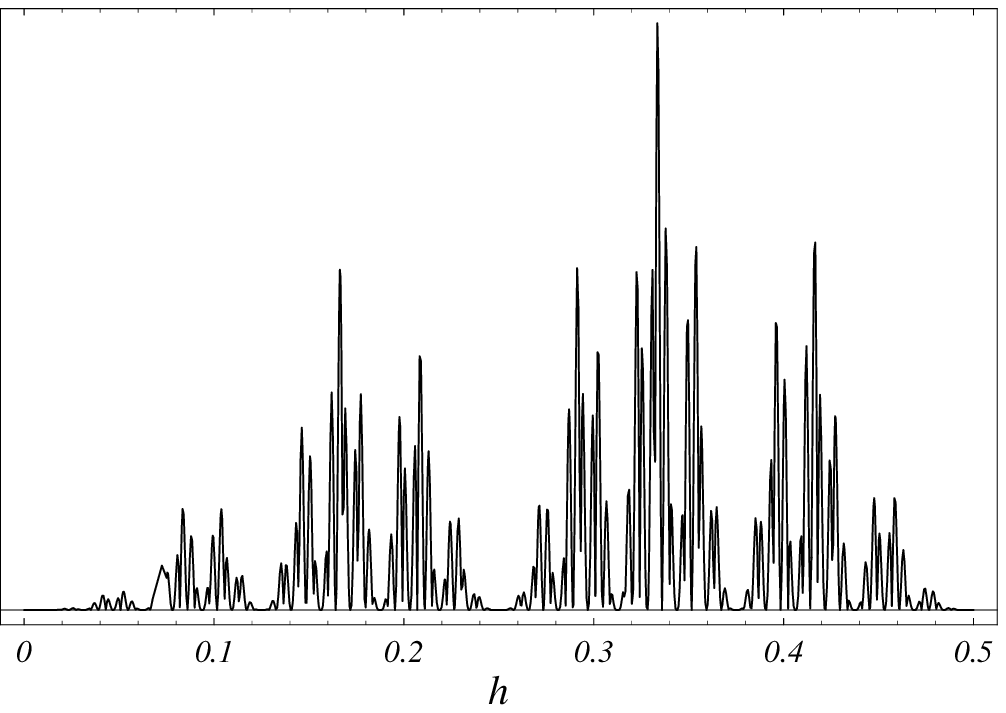,width=6.5cm}}
\caption{Diffraction image of the 2D Thue-Morse product structure (top). Cut along the
line $k=\frac{1}{3}$ (bottom). The system size for the numerical calculations
was $512 \times 512$.}\label{tm2ddiff}
\end{figure}

The resulting structure is a simple Cartesian product of Thue-Morse chains and can be
treated using the method indicated in \cite[Sec.\ 4]{BH}. The diffracted intensity is a product
measure
$\widehat{\gamma}_{\omega}^{2\text{D}}=\widehat{\gamma}_{\omega^{(1)}}^{\text{TM}} 
\cdot \widehat{\gamma}_{\omega^{(2)}}^{\text{TM}}$
of the corresponding one-dimensional intensities 
$\widehat{\gamma}_{\omega^{(i)}}^{\text{TM}}$ (compare \cite{GL}). We assign 
scattering strenghts of $+1$ (full
circles) and $-1$ (open circles) to the atoms. Taking values $1$ and $0$ would just
lead to an additional, trivial Bragg part on $\mathbb Z^2$. The remaining diffracted 
intensity is a
product of singular continuous measures \cite{Quef,GL} and thus itself
purely singular continuous (Fig.\ \ref{tm2ddiff}). We have to emphasize that this
singular continuous behaviour is by no means recognizable from the shape of the peaks 
and the decrease of their wings. The most likely fit is given by Gaussians, but this
is also the case for true Bragg peaks.

The explicit scaling of the peaks with the system size can be obtained by 
applying the formulae of \cite[Sec.\ 4.3]{GL} to our product structure.

\section{The Ising lattice gas and its symmetry}

It's worth taking a glance at the well-known 2D Ising model where, in contrast to the
previous examples, we have some interaction between the atoms. In the lattice gas
formulation, we place atoms and holes (scattering strengths 1 and 0) on the square
lattice subjected to coupling constants $K_i=J_i/k_B T > 0$ in $x$- and $y$-direction. 
There is a phase transition at $(\sinh (2 K_1) \sinh ( 2 K_2))^{-1} =
1$. One can show \cite{BH} that the pure point part of the diffraction spectrum is
fourfold symmetric for all $K_i$ and has the form
\begin{equation}
(\widehat{\gamma}_{\omega})_{pp}= 
\begin{cases}
\frac{1}{4} \sum_{\boldsymbol k \in \mathbb Z^2} \delta_{\boldsymbol k},& \quad T \geq
T_c \\
\rho^2 \sum_{\boldsymbol k \in \mathbb Z^2} \delta_{\boldsymbol k},& \quad T < T_c,
\end{cases}
\end{equation}
where $\rho$ is the ensemble average of the number of scatterers per unit volume, $1/2
\leq \rho \leq 1$.

\begin{figure}[ht]
\centering \epsfig{file=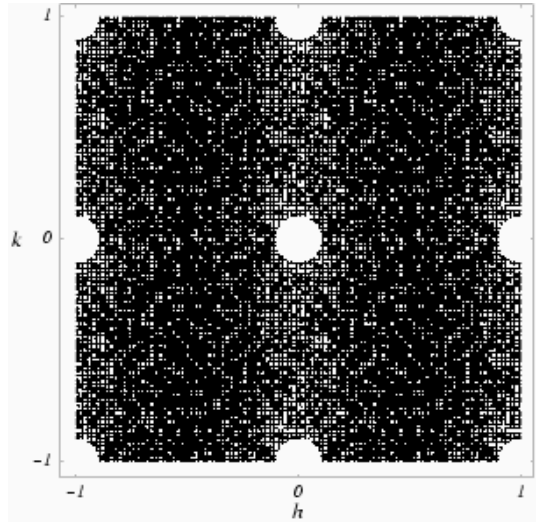,width=7cm}
\caption{Ising lattice gas above $T_c$ with $K_1 = 0.35$ and $K_2 = 0.1$.}
\end{figure}  

The correct (two-fold) symmetry is visible only from the diffuse part of the spectrum,
the Bragg part alone is misleading. Using the
asymptotic behaviour of the correlation functions, the background can be shown to be
an absolutely continuous measure \cite{BH}, which can be represented by a continuous
function. The background concentrates around the Bragg peaks, as expected for an
attractive interaction.

\section{The ice model}

\begin{figure}[ht]
\centering\epsfig{file=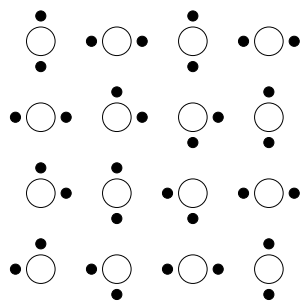,width=3.9cm}
\caption{Typical square ice configuration}
\label{icestruct}
\end{figure}

Square ice is another example for which the relevant structural information of the
diffraction image is only contained 
in the diffuse background. In this model, oxygen atoms are sitting on the vertices of
a square lattice, see Fig.\ \ref{icestruct}. The hydrogens are placed on the edges with 
a distance of $1/3$ to
the next O according to the following ice (or Bernal-Fowler) rules (\cite{Petr} and
references therein):
\begin{enumerate}
\item There are two hydrogens adjacent to each oxygen.
\item There is only one hydrogen per bond.
\end{enumerate}

By using the equivalence to a special case of the six-vertex model, Lieb \cite{Lieb} 
solved the ice model and calculated its residual entropy to be $\frac{3}{2} \log
\frac{4}{3}$. Although calorimetric measurements on three-dimensional ice
confirmed the predicted value of the entropy very well, it was almost 
impossible to detect the hydrogen disorder in X-ray scattering experiments. Neutron
studies are more promising, compare \cite{Kuhs} where a highly disordered hydrogen
structure was shown in deuterated cubic ice Ic. 

Apart from the weak scattering power of the hydrogens, another more fundamental
problem arises. Consider the case of a Bernoulli system where we ignore the
first ice rule, so that the hydrogens are distributed with equal probability 
independently of the others. 

By assigning scattering strengths $h_{\text{O}}$ and $h_{\text{H}}$ to the atoms, a straightforward
calculation  yields the autocorrelation of the Bernoulli model:
\begin{multline}
\gamma_{\omega}=\sum_{\boldsymbol z\in\mathbb Z^2} \Bigl( h_{\text{O}}^2 + 
h_{\text{H}}^2 + \frac{h_{\text{H}}^2}{2} \sum_{x,y =\pm \frac{1}{3}} \delta_{(x,y)} \\
+ \Bigl( h_{\text{O}} h_{\text{H}} +
\frac{h_{\text{H}}^2}{4}\Bigr)\sum_{x=\pm\frac{1}{3}}
\left( \delta_{(x,0)}+\delta_{(0,x)}\right) 
\Bigr) \ast \delta_{\boldsymbol z}  \\
+h_{\text{H}}^2\Bigl( \delta_0 -\frac{1}{4} \sum_{x=\pm\frac{1}{3}} 
\left( \delta_{(x,0)}+\delta_{(0,x)}\right)\Bigr).
\end{multline}

Thus, the Bragg part is given by $(\boldsymbol k = (k_1,k_2))$
\begin{equation}
\label{eq:bragg}
\bigl( \widehat{\gamma}_{\omega} \bigr)_{pp} = 
\sum_{\boldsymbol k \in \mathbb Z^2} \biggl( h_{\text{O}}+h_{\text{H}} \Bigl( \cos \frac{2 \pi k_1}{3}+
\cos \frac{2 \pi k_2}{3} \Bigr)\biggr)^2 \delta_{\boldsymbol k},
\end{equation}
and the absolutely continuous background can be represented by the continuous function
\begin{equation}
\label{eq:back}
g({\boldsymbol k}) = h_{\text{H}}^2\biggl( \sin^2 \frac{\pi k_1}{3}+\sin^2 
\frac{\pi k_2}{3}\biggr).
\end{equation}

But due to the same averaged occupation probabilities of the hydrogens in the ice and the
Bernoulli model, their pure point parts are indistinguishable. In other words, Eq.\ 
(\ref{eq:bragg}) is valid for the case of square ice, too. 
The background is
different, of course, but very weak, in particular because, in reality,
$h_{\text{O}} \gg h_{\text{H}}$.

\begin{figure}[ht]
\centering\epsfig{file=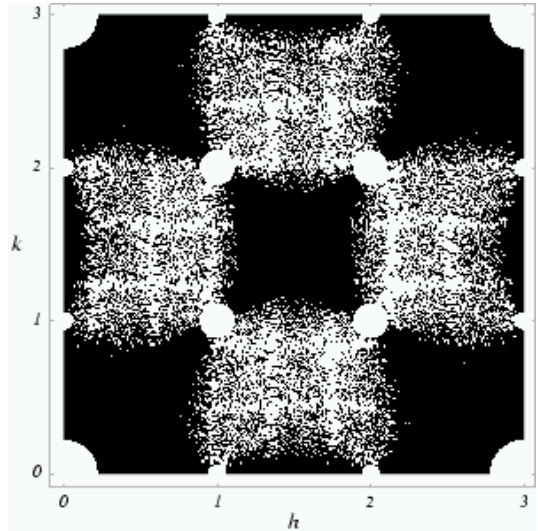,width=7cm}
\caption{Elementary cell of the diffraction pattern for square ice ($h_{\text{O}} = 0$ 
and $h_{\text{H}} = 1$). Only the weak
(even for vanishing $h_{\text{O}}$!) background differs from the Bernoulli system,
compare Eq.\ (\ref{eq:back}).}
\label{icediff}
\end{figure}

Fig.\ \ref{icediff} displays the diffraction image of square ice. To focus on the
contribution of the hydrogens, we set $h_{\text{O}} = 0$. Positive $h_{\text{O}}$ would 
result in an increase of intensity of the Bragg
peaks, except for the case of $h_{\text{O}}= h_{\text{H}} $ where we have extinction at 
$(k_1,k_2)=(1,1)$, (1,2),
(2,1) and (2,2), according to Eq.\ (\ref{eq:bragg}). The continuous background,
however, depends only on $h_{\text{H}}$, as in the Bernoulli case.

We have no explicit formula for the background in the ice model, since the correlation
functions are not known, but we can calculate the
diffraction image numerically (FFT) from a pattern of a Monte Carlo simulation, see 
\cite[Ch.\ 7]{NB} for the method. 
The decay of the correlations suggests that the background is an absolutely continuous 
measure and there is no singular continuous component present. Note that the
interpretation of the diffuse part is more involved here -- it needs to include
multi-particle effective interactions.

\end{document}